\documentclass{article}
\usepackage{spconf,amsmath,graphicx}
\usepackage{graphicx}
\usepackage{fixltx2e}
\usepackage{subcaption}
\usepackage{color}
\usepackage{amsfonts}
\usepackage[caption=false,font=footnotesize]{subfig}

\definecolor{mygreen}{rgb}{.1,.6,.1}
\definecolor{orange}{rgb}{1,.5,0}

\title{A SYSTEM FOR HIGH PRECISION GlASS-TO-GLASS DELAY MEASUREMENTS IN VIDEO COMMUNICATION}
\name{Christoph Bachhuber and Eckehard Steinbach}
\address{Technical University of Munich\\ Chair of Media Technology\\  
Munich, Germany}
\begin{document}
%
\maketitle
\begin{abstract}
Ultra low delay video transmission is becoming increasingly important. Video-based applications with ultra low delay requirements range from teleoperation scenarios such as controlling drones or telesurgery to autonomous control of dynamic processes using computer vision algorithms applied on real-time video. To evaluate the performance of the video transmission chain in such systems, it is important to be able to precisely measure the glass-to-glass (G2G) delay of the transmitted video. In this paper\footnote{Copyright \copyright 2016 IEEE, article accpeted for publication by IEEE. Personal use of this material is permitted. However, permission to use this material for any other purposes must be obtained from the IEEE by sending an email to pubs-permissions@ieee.org. DOI: 10.1109/ICIP.2016.7532735}, we present a low-complexity system that takes a series of pairwise independent measurements of G2G delay and derives performance metrics such as mean delay or minimum delay etc. from the data. The precision is in the sub-millisecond range, mainly limited by the sampling rate of the measurement system. In our implementation, we achieve a G2G measurement precision of 0.5 milliseconds with a sampling rate of 2kHz.
\end{abstract}

\section{Introduction}
\label{sec:intro}

With the advent of 5G networks \cite{boccardi2014five} and the prospects of the tactile internet \cite{fettweis2014tactile} End-to-End (E2E) delays of 1 millisecond are requested for communication systems of the future. These ultra low delay systems enable applications such as networked control for fast assembly robots, highly dynamic teleoperation \cite{lum2008teleoperation} in virtual or augmented reality\cite{nielsen2007ecological}, car-to-X communication \cite{david2010car, festag2008car} to improve safety and efficiency in transport, and many more. 

In all these scenarios, ultra low delay video transmission is an important component. Therefore, we need a precise measurement of the E2E video delay. For video transmission systems, which presents the video to a user on a display, this is called Glass-to-Glass (G2G) delay. It describes the time from when the photons of a visible event pass through the lens of a camera until the corresponding photons of the event shown on a display pass through the display glass.

The G2G measurements are preferably non-intrusive, such that they can be applied to a wide range of systems. Furthermore, a video camera typically has a fixed refresh rate, producing new images in constant time intervals. Real world events are virtually never synchronized to the camera frame capture time instances. To make realtime measurements, real-world events have to be triggered. Because of this non-deterministic G2G delay values are obtained. By repeating the measurement process several times, a distribution of delay values is obtained.

Measuring partial delays such as the processing delay on a camera or the encoding latency is a standard task in system design. For both, the signal propagation time through the circuit has to be measured. But there are few approaches available to measure the G2G latency of the more complex system of an entire video transmission chain. This measurement also comprises delays from data transmission between processing blocks and the synchronization effects between blocks operating at fixed rate.

\subsection{Related Work}
\label{ssec:relw}
Several methods to measure G2G delay in video transmission have previously been proposed. An overview of their system characteristics is given in Table \ref{tab:comp}. 

\begin{table}
	\begin{center}
	{\renewcommand{\arraystretch}{1}
	\begin{tabular}{p{2.18cm}|p{0.8cm}|p{0.82cm}|p{0.85cm}|p{0.7cm}|p{0.85cm}}
		Author & Auto- matic & Non- Intrusive & De- correlated & Cost & Pre- cision \\
		\hline
		Hill/MC \cite{hill2009measuring, maccormick2013video} 
		& \textcolor{red}{no} & \textcolor{mygreen}{yes} & \textcolor{red}{no} & \textcolor{orange}{med} & \textcolor{red}{low} \\
		Jacobs \cite{jacobs1997managing}
		& \textcolor{red}{no} & \textcolor{mygreen}{yes} & \textcolor{red}{no} & \textcolor{orange}{med} & \textcolor{mygreen}{high} \\
		Sielhorst \cite{sielhorst2007measurement} 
		& \textcolor{mygreen}{yes} & \textcolor{mygreen}{yes} & \textcolor{red}{no} & \textcolor{orange}{med} & \textcolor{red}{low} \\
		Boyaci \cite{boyaci2009vdelay}
		& \textcolor{mygreen}{yes} & \textcolor{red}{no} & \textcolor{red}{no} & \textcolor{mygreen}{none} & \textcolor{red}{low}\\
		Jansen \cite{jansen2013user}
		& \textcolor{mygreen}{yes} & \textcolor{mygreen}{yes} & \textcolor{red}{no} & \textcolor{red}{high} & \textcolor{red}{low} \\
		Our method
		& \textcolor{mygreen}{\textcolor{mygreen}{yes}} & \textcolor{mygreen}{yes} & \textcolor{mygreen}{yes} & \textcolor{mygreen}{low} & \textcolor{mygreen}{high} \\
	\end{tabular}}
	\end{center}
	\caption{Comparison of delay measurement methods. Justification of the classification is given in Section \ref{ssec:relw}. Our method is presented in Section \ref{sec:sd}.}
	\label{tab:comp}
\end{table}

The approaches in \cite{hill2009measuring, maccormick2013video} rely on the presentation of a running clock, for example on a computer screen. This clock is filmed, the video of it transmitted and displayed by the video transmission system under test. Another camera films both the real clock and the clock displayed by the video transmission system. By comparing the clock states in the resulting image, the G2G delay can be obtained. These methods suffer from many issues: without image processing algorithms, the calculation of the delay has to be done manually by reading the numbers from the final image. For the measurement system, one has to purchase an additional camera to record the entire scene. Further, the achievable precision is low because the monitor displaying the running clock and the second camera are refreshed at their individual frame rates, e.g. $f_{\rm Dis} = f_{\rm Cam} =60$Hz.


Jacobs et al. \cite{jacobs1997managing} set the basis for our system: the authors use a blinking light-emitting diode (LED) in the field of view of the camera as signal generator and tape a photoelectric sensor to where the LED is shown on the display. The LED triggers an oscilloscope which also records the signals from the photoelectric sensor. This allows them to manually extract the G2G delay of individual samples. The problem is that this method is not automated on a simple circuitry and therefore requires high effort and expensive equipment. 

Sielhorst et al. \cite{sielhorst2007measurement} propose a system that comprises moving LEDs. From the position difference of the LEDs in the actual world and on the video, the delay is automatically computed by employing a computer vision algorithm. This method does not include the exposure delay of the camera since the source continuously creates events (new translation positions). 
Furthermore, they use a recording rate of the measurement camera of at most 200 Hz. This introduces an average imprecision of 5 milliseconds.

Boyaci et al. \cite{boyaci2009vdelay} measure the capture-to-display latency between a caller and a callee in a video conferencing application. They embed timing information in the form of an EAN-8 barcode in the recorded frames. This information is decoded on the callee PC and compared to the internal clock in software. The method is constrained to desktop computers, since it is intrusive and requires custom software to be executed on the caller and callee machines. The authors assume synchronized clocks and take no further analyses or measures to ensure synchronization. Finally, the method does not include the delay introduced by the graphics buffer and the display, since the timestamp is compared to the current time immediately after decoding.

Jansen et al. \cite{jansen2013user} utilize QR codes to mark time. A measurement system feeds QR codes from a display to the camera of the system under test, from which the video is displayed and again recorded by the measurement system. The measurement system decodes the QR code and computes the G2G delay. The problem is that a camera is not a time-precise recording tool. Further, a computer or laptop and a camera have to be used as measurement system, which constitutes one of the most expensive options here.

\begin{figure}
  \centering
  \includegraphics[width=0.45\textwidth]{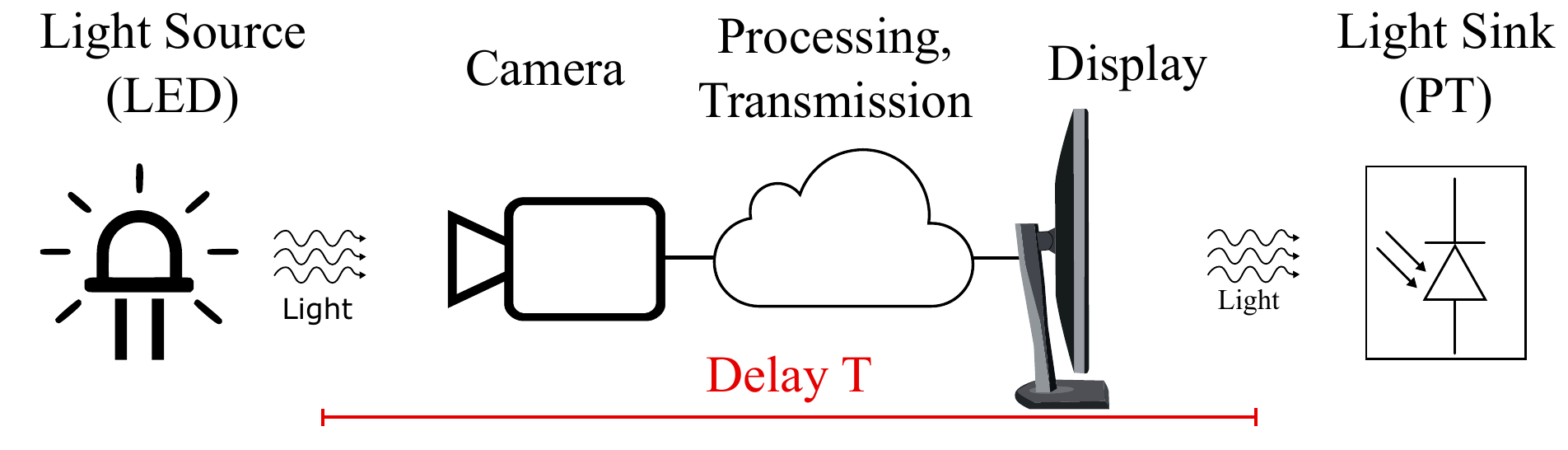}
  \caption{Delay measurement principle}
  \label{fig:dmp}
\end{figure}

\subsection{Contribution}
We propose a G2G delay measurement system that unifies most of the benefits of the existing systems as shown in Table\ref{tab:comp}. It is an advancement of Jacobs' \cite{jacobs1997managing} system and comprises an LED as light source and a phototransistor (PT) as light detector. The actual LED can cover only a small area of the video image to not bias the coding process. The analysis of the data is not done manually with an oscilloscope, but automatically with a microcontroller board. We propose a theoretical model for G2G delay and relate initial measurements obtained with the new system to it.

The remainder of this paper is organized as follows: Section \ref{sec:sd} describes the system principle, the hardware and software implementation and a theoretical model for delay. Section \ref{sec:res} presents and discusses results obtained with the measurement system. Section \ref{sec:con} summarizes the results and gives an outlook to future work in this field.

\section{System Description}
\label{sec:sd}
\subsection{Concept and Realization}
\label{ssec:pr}

The G2G delay measurement process is based upon the idea that the video transmission system delays the propagation of light, as depicted in Figure \ref{fig:dmp}. An initially disabled light source is put in the field of view of the camera. After enabling the light source, the video transmission system requires the G2G delay $T$ to transmit this information to the display, which is picked up by the light sink. The proposed approach assumes an ideal system without any reaction delay within the light source and sink and with no noise.

\begin{figure}
  \centering
	\begin{picture}(221,150)
		\put(0,0){\includegraphics[width=0.45\textwidth]{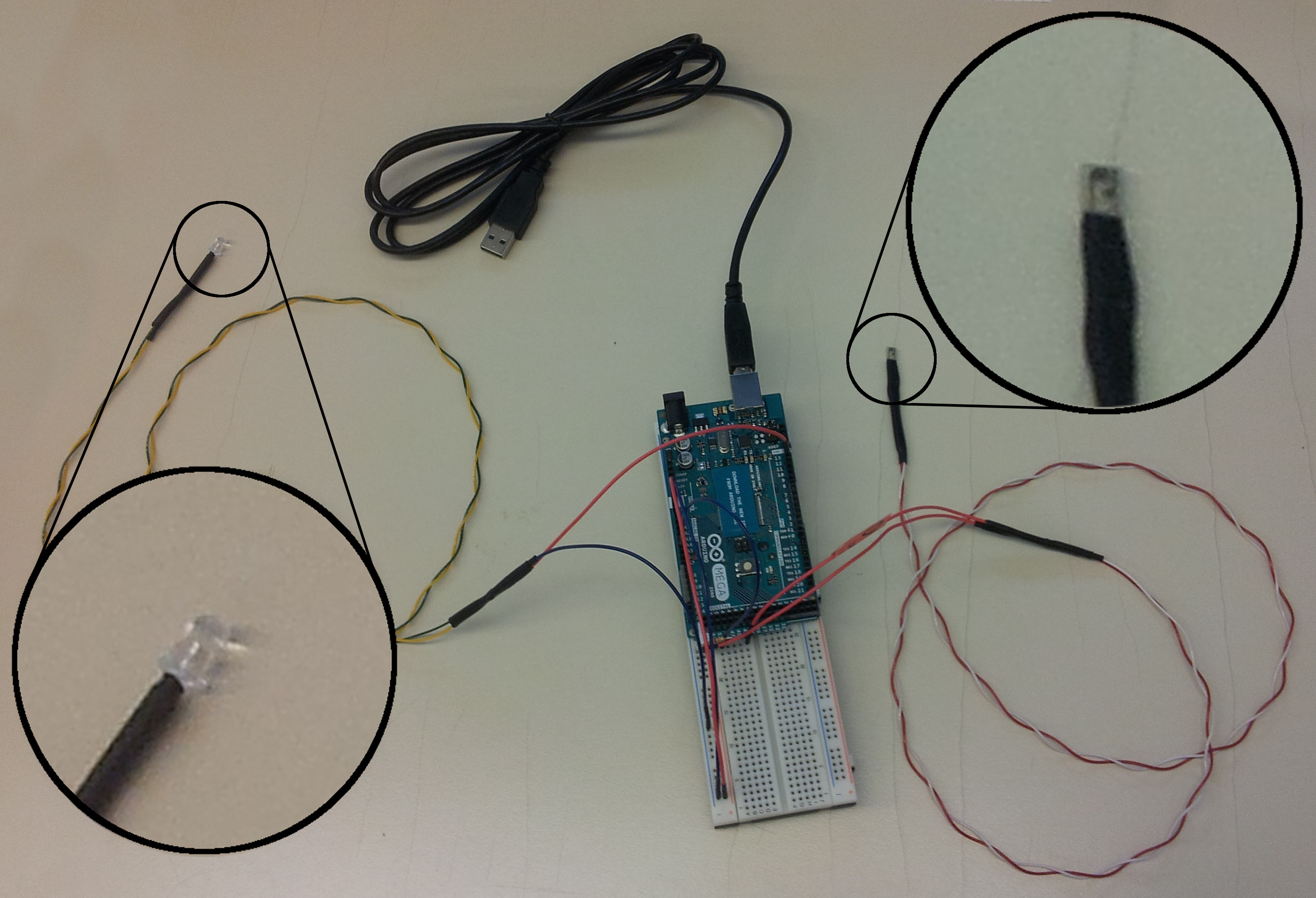}}
		\put(23,54){\Large LED}
		\put(181,132){\Large PT}
		\put(70,32){\large Arduino\textsuperscript{\textregistered}}
		\put(98,112){\large USB}
	\end{picture}
  \caption{Prototype}
  \label{fig:real}
\end{figure}
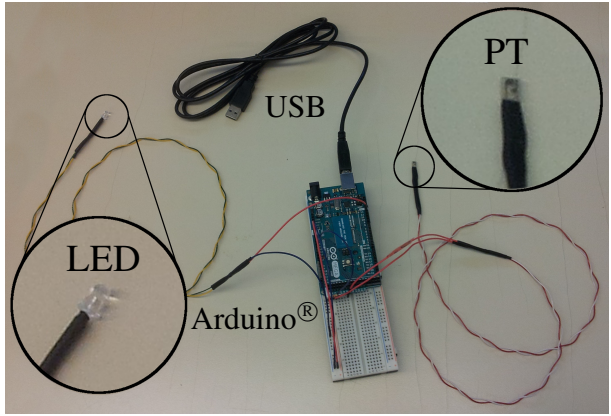

We created a prototype with an Arduino \textsuperscript{\textregistered} Uno. It is depicted in Figure \ref{fig:real}. It can be connected to a PC using USB or to mobile devices using bluetooth. An LED acts as light source in the field of view of the camera. In LEDs, the time between the start of an electrical current pulse and the start of emission of photons is typically below one microsecond. Since our measurements are in the order of milliseconds, the delay from the LED is negligible. The light sink is a phototransistor (PT) which has a rise and fall time of 10 microseconds, which is also small compared to the G2G delay we want to be able to measure. To suppress noise we are using the detection algorithm proposed in Section \ref{ssec:red}.

\subsection{Signal Processing} 
\label{ssec:red}
The voltage dropping over the PT is sampled at 2kHz in our prototype. The resolution of the voltage is 10 bit, resulting in 1024 brightness levels. To extract the time at which the event appears on the display, the sample data undergoes a two-step processing: first, a maximum smoothing filter and second a rising edge detection algorithm are applied (both steps are described below). The algorithm has been validated by comparing the resulting G2G delays with manually read values from an oscilloscope which is connected to the LED and PT.

The maximum smoothing is required to suppress wrong detections caused by pulse width modulation (PWM) of LCD display backlight or short light pulses in CRT and plasma monitors. The filter has two tasks: smooth the signal from unwanted waves and let the resulting signal increase immediately if the input signal increases. This is solved by the maximum filter with length $k$. For every new raw sample $a_i$, the maximum

\[
b_i = \max_{\max(0,i-k) \leq j \leq i} (a_j)
\]

of itself and the previous $k$ samples is stored in the processed value $b_i$. 

To automatically find the sample at which a consistent increase of the sample values is initiated, we apply a rising edge detection based on slope thresholding on the processed samples $b_i$. An increase of a cumulative 20 brightness levels over the duration of 3 subsequent samples or the same increase within one sample to the next triggers the flag that the picture of the lit up LED can now be seen on the display. These parameters make the algorithm robust against noise from external lighting and panel refresh on one hand. On the other hand, it enables us to reliably recognize the lighting up of the LED in typical measurement environments without further precautions.

With constant inter-measurement intervals, a measurement sequence of a simple Camera to PC setup exhibits strong correlations between measurement samples, considerably reducing their significance. This is because of the constantly changing phase shifts between the sampling processes in the camera and display. To avoid these correlations, we use random inter-measurement intervals.

\subsection{Delay Distribution}
\label{ssec:sampling}
To explain the G2G measurements obtained with the proposed system, we model the G2G delay distribution of a simple raw video transmission system consisting of a camera, a PC and a display. We first define three partial delays: the camera sampling delay $p_{\rm Cam}(t) \sim \mathcal{U}(t_{\rm min},f_{\rm Cam}^{-1}+t_{\rm min})$ contributed by the camera sampling is uniformly distributed because the turn-on time of the LED is independent of the frame period $f_{\rm Cam}^{-1}$ of the camera. The LED has to light up at least $t_{\rm min}$ before the end of a frame period $f_{\rm Cam}^{-1}$ to be part of the current frame. This frame is read out of the sensor and transmitted at the end of the current frame period, leading to a delay in the interval $[t_{\min},f_{\rm Cam}^{-1}]$. If the LED turns on later than that during the current frame period, the light-up information is transmitted at the end of the next frame period, causing a delay in $]f_{\rm Cam}^{-1},f_{\rm Cam}^{-1} + t_{\min}]$. These two possibilities together form the uniform distribution of $p_{\rm Cam}(t)$ as seen in the beginning of this paragraph. The occurrence of the second possibility has two reasons: first, it either lights up so late during the exposure that the corresponding relatively dark depiction on the display will not trigger the rising edge detection. Second, the LED can light up during a frame period after the exposure has ended. The minimum exposure required for triggering and the difference between a frame period $f_{\rm Cam}^{-1}$ and the exposure time add up to $t_{\rm min}$.


The display refresh also contributes a uniform delay $p_{\rm Ref}(t) \sim \mathcal{U}(0,f_{\rm Dis}^{-1})$, upper bounded by the inverse of the display refresh rate $f_{\rm Dis}^{-1}$. This delay is uniformly distributed because the display refreshes independently of when the computer fills the graphics buffer. 

All remaining parts like the processing in the camera, PC and display and the interface delays are modeled to be deterministic and are thus represented by one variable $p_{\rm Proc}(t) \sim \delta(t_{\rm Proc})$. In reality, there will be deviations from the ideal deterministic delay for example because we do not use a real-time operating system.

Since the G2G-delay $T$ is the sum of these three mutually independent delays, the corresponding probability distribution 

\[
	T \sim P(t) = p_{\rm Cam}(t) * p_{\rm Proc}(t) * p_{\rm Ref}(t)
\]

is the convolution of them. The G2G delay distribution approximates an isosceles trapezoid shape that is centered around the mean $t_{\rm Proc} + t_{\min} + \frac{1}{2 f_{\rm Cam}} + \frac{1}{2 f_{\rm Dis}}$ with minimum delay $t_{\rm Proc} +t_{\min}$ and maximum latency $t_{\rm Proc} + t_{\min} + \frac{1}{f_{\rm Cam}} + \frac{1}{f_{\rm Dis}}$. In real measurements, the non-deterministic processing delay will smoothen the nooks of the shape. 

\section{Measurements}
\label{sec:res}
We present measurements conducted with our prototype described in Section \ref{sec:sd}. The video transmission system is a Fedora 20 PC with an AlliedVision Guppy PRO F-031C IEEE 1394 camera and a Samsung 2233BW monitor at $f_{\rm Dis}=60$Hz. We parametrized the camera such that the exposure time is, with a negligibly small difference below the millisecond order, equal to the frame period. As displaying software, we use coriander 2.0.2. 

\begin{figure}[t!]
	\centering
	\begin{subfigure}[b]{0.45\textwidth}
  \includegraphics[width=\textwidth]{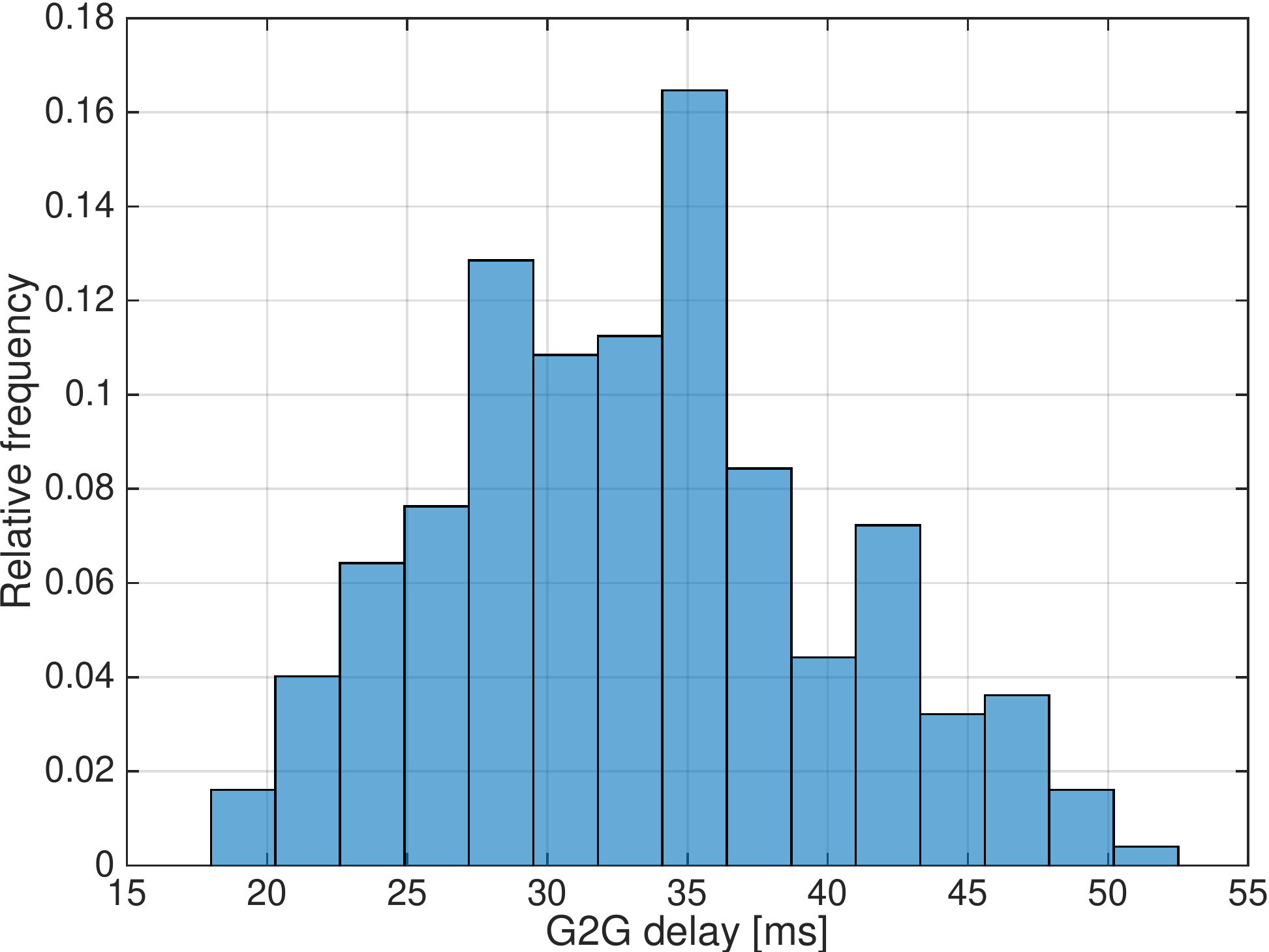}
  \caption{G2G delay measurement distribution for 50Hz camera frame rate}
  \label{fig:Adel}
	\end{subfigure}
	\begin{subfigure}[b]{0.45\textwidth}
  \includegraphics[width=\textwidth]{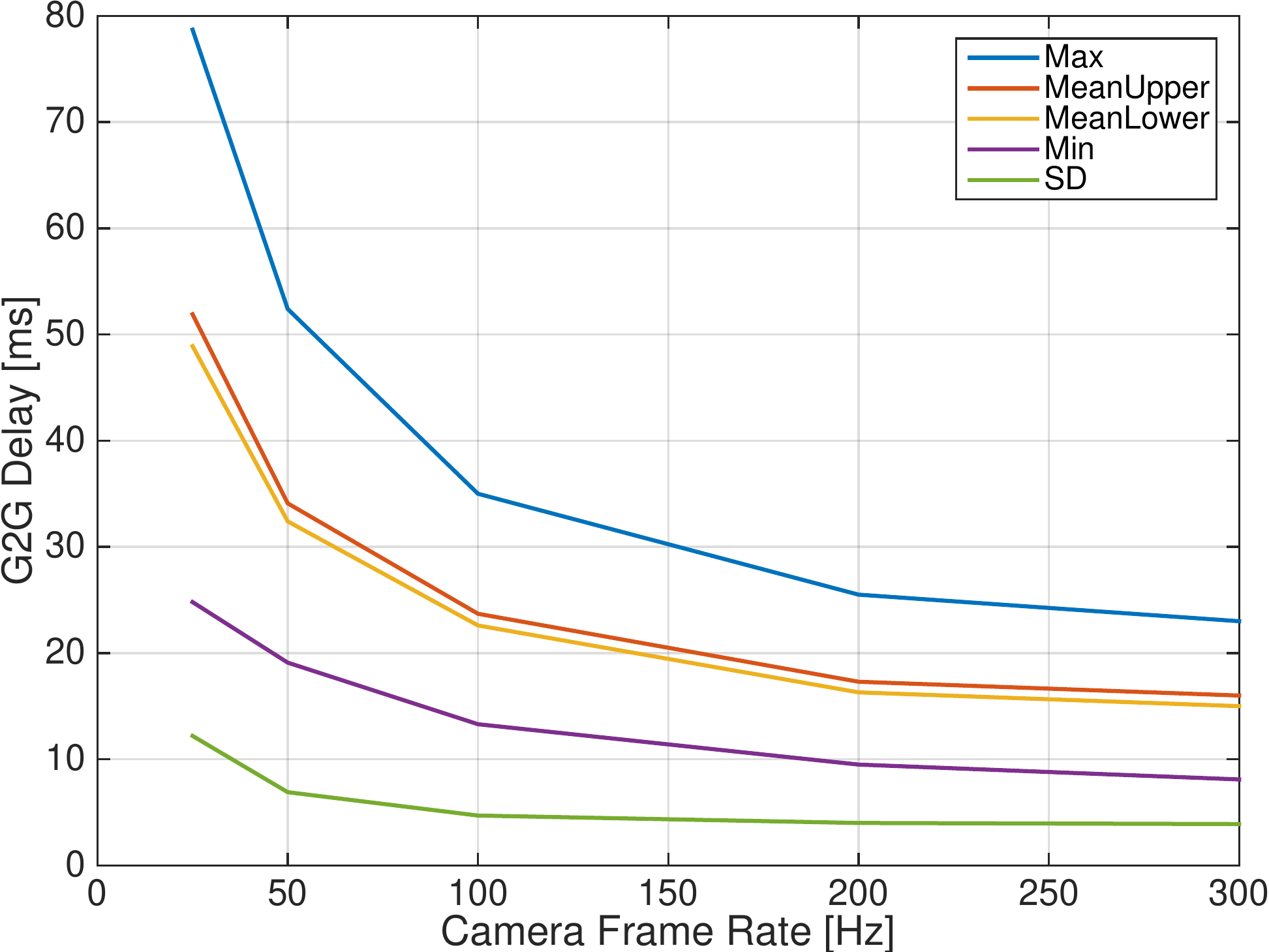}
  \caption{G2G delay measurement distribution characteristics for different frame rates of the camera}
  \label{fig:gup}
	\end{subfigure}
	\caption{Measurements}
	\label{fig:sig}
\end{figure}

The G2G delay distribution of 250 measurements with $f_{\rm Cam} = 50$Hz is shown in Figure \ref{fig:Adel}. The delay is at minimum 19.1ms$=t_{\rm Proc} + t_{\min}$. The sum elements can not be distinguished using the data produced by the proposed measurement system. With this minimum delay, it takes at minimum 19.1ms from an event taking place until it is shown on the display. This can also be thought of as the best case measurement. The opposite, the maximum delay is 52.4ms$=t_{\rm Proc} + t_{\min} + \frac{1}{f_{\rm Cam}} + \frac{1}{f_{\rm Dis}}$, representing the worst case delay from the event until the display of it. The 95\% confidence interval from fitting a Student's $t$-distribution to the histogram in Figure \ref{fig:Adel} for the mean ranges from 32.4ms to 34.1ms. The standard deviation is 6.9ms. The histogram in Figure \ref{fig:Adel} also confirms the assumptions from Section \ref{ssec:sampling}: it approximates an isosceles trapezoid and has a width of $52.4{\rm ms}-19.1{\rm ms}=33.3$ms. This is a few milliseconds smaller than $\frac{1}{f_{\rm Cam}} + \frac{1}{f_{\rm Dis}} \approx 20 {\rm ms} + 16.7{\rm ms} = 36.7{\rm ms}$ because the ideal worst and best case delays are so improbable that they did not occur in this series of measurements. Performing more measurements reduces the difference in width between theory and practical measurements. But with an increasing number of measurements, the difference only approximates zero, but does not perfectly equal it. This is why we did not perform more measurements here.

In Figure \ref{fig:gup}, we plot maximum G2G delay, the bounds for the 95\% confidence interval for estimating the mean, the minimum delay and the standard deviation of the delay as a function of the frame rate of the camera. For every frame rate setting, 250 G2G measurements have been performed. The statistics of the measurements in Figure \ref{fig:Adel} can be seen at 50Hz in Figure \ref{fig:gup}. All statistics are monotonically decreasing with ascending frame rates. This is because $f_{\rm Cam}^{-1}$, influencing the camera sampling delay, gets smaller. $t_{\min}$ decreases because with ascending frame rates, we increase the gain of the camera sensor, which allows the LED to be turned on later during exposure and still be detected by the PT. 
The 95\% confidence interval for the mean estimation lies between the curves MeanUpper and MeanLower. The delay distributions of the different frame rates resemble the distribution in Figure \ref{fig:Adel}, thus providing no further insight and are therefore not depicted. 

The triple \textbf{(minimum delay / mean delay / maximum delay)} sufficiently describes the G2G delay characteristics of a system, so this is the metric we report. For the $f_{\rm Cam}=25$Hz and $f_{\rm Cam}=300$Hz camera frame rates, these are $(24.8/50.4/78.7)$ms and $(8.1/15.5/23)$ms, respectively. For $f_{\rm Cam}=25$Hz, the width of the histogram, which is the difference between minimum and maximum delay, is 53.9ms, approximating $\frac{1}{f_{\rm Cam}} + \frac{1}{f_{\rm Dis}}\approx 40{\rm ms} + 16.7{\rm ms} = 56.7{\rm ms}$. An analog approximation holds over all measured camera frequencies, which again confirms the model from Section \ref{ssec:sampling}.

\section{Conclusions}
\label{sec:con}
We proposed an inexpensive, automatic and highly precise G2G delay measurement system. It unifies advantages of previously proposed implementations and can be used to independently assess bigger, more complex video transmission systems. Furthermore, we briefly discussed the origins of delay in video transmission and showed that the measurements fit to the proposed model.

\bibliographystyle{IEEEbib}
\bibliography{latex8}

\end{document}